\documentclass[sigconf]{acmart}
\usepackage{amsmath}
\usepackage{algorithm}
\usepackage{algpseudocode}

\usepackage{subfigure}
\usepackage{graphicx}
\usepackage[utf8]{inputenc}
\usepackage[neverdecrease]{paralist}
\usepackage{color}
\usepackage[normalem]{ulem}
\newcommand{\eatme}[1]{ }
\makeatletter
\usepackage{paralist}
\usepackage{xspace}

\usepackage{booktabs} 

\editor{Jennifer B. Sartor}
\editor{Theo D'Hondt}
\editor{Wolfgang De Meuter}

\begin{document}

\copyrightyear{2018} 
\acmYear{2018} 
\setcopyright{acmcopyright}
\acmConference[ICS '18]{2018 International Conference on Supercomputing}{June 12--15, 2018}{Beijing, China}
\acmBooktitle{ICS '18: International Conference on Supercomputing, June 12--15, 2018, Beijing, China}
\acmPrice{15.00}
\acmDOI{10.1145/3205289.3205304}
\acmISBN{978-1-4503-5783-8/18/06}

\title{Dynamic Load Balancing for Compressible Multiphase Turbulence
}

\author{Keke Zhai}
\affiliation{
  \institution{CISE, University of Florida}
  }
\email{zhaikeke@ufl.edu}

\author{Tania Banerjee}
\affiliation{
  \institution{CISE, University of Florida}
  }
\email{tmishra@ufl.edu}

\author{David Zwick}
\affiliation{
  \institution{MAE, University of Florida}
  }
\email{dpzwick@ufl.edu}
\vspace{-0.6cm}

\author{Jason Hackl}
\affiliation{
  \institution{MAE, University of Florida}
  }
\email{jason.hackl@ufl.edu}
\vspace{-0.2cm}

\author{Sanjay Ranka}
\affiliation{
  \institution{CISE, University of Florida}
  }
\email{ranka@ufl.edu}
\vspace{-0.2cm}

\begin{abstract}
CMT-nek is a new scientific application for performing high fidelity predictive simulations of particle laden explosively dispersed turbulent flows. CMT-nek involves detailed simulations, is compute intensive and is targeted to be deployed on exascale platforms. The moving particles are the main source of load imbalance as the application is executed on parallel processors. In a demonstration problem, all the particles are initially  in a closed container until a detonation occurs and  the particles move apart. If all processors get an equal share of the fluid domain, then only some of the processors get sections of the domain that are initially laden with particles, leading to disparate load on the processors. In order to eliminate load imbalance in different processors and to speedup the makespan, we present different load balancing algorithms for CMT-nek on large scale multi-core platforms consisting of hundred of thousands of cores. The detailed process of the load balancing algorithms are presented. The performance of the different load balancing algorithms are compared and the associated overheads are analyzed. Evaluations on the application with and without load balancing are conducted and these show that with load balancing, simulation time becomes faster by a factor of up to $9.97$. 
\end{abstract}
\vspace{-0.6cm}

\maketitle

\vspace{-0.3cm}
\section{Introduction}
Flows with compressible multiphase turbulence (CMT) are difficult to accurately predict and simulate because of complexities of the actual physical processes involved. CMT-nek~\cite{Banerjee-7839682} is a new scientific application that solves the compressible Navier-Stokes equations for multiphase flows. This application is designed to perform high fidelity, predictive simulations of particle laden explosively dispersed turbulent flows under conditions of extreme pressure and temperature. The actual physical processes underlying explosive dispersal are complex and cover a very wide range of temporal and spatial scales. Therefore, the simulations
require enormous computing power and CMT-nek is expected to be deployed in petascale and exascale supercomputers.

CMT-nek leverages the code setup and data structures used in Nek5000 which is an open source spectral element based computational fluid dynamics code developed at Argonne National Laboratory for simulating unsteady incompressible fluid flow with thermal and passive scalar transport ~\cite{c19, nek5000, dfm02}. Nek5000 is a highly scalable code, with demonstrated strong scaling to over a million MPI ranks on ALCF BG/Q Mira. Nek5000 is, however, limited to low speed flows by its formulation and discretization. Further, Nek5000 does not simulate particle laden flows. As a simulation workhorse, CMT-nek is expected to facilitate fundamental breakthroughs and development of better (physics-informed) models and closures for compressible multiphase turbulence.

Nek5000 uses a static load balancing strategy, where the domain is first partitioned into spectral elements. The domain partitioner uses a recursive spectral bisection algorithm \cite{hendrickson1993multidimensional} and arranges the resulting spectral elements in a one-dimensional array preserving spatial locality. The one-dimensional array of spectral elements is then equally divided among the processors. CMT-nek, on the other hand, has a dynamic load balancing strategy. Firstly, CMT-nek estimates the ratio of computational load between particles and fluid. Then, the total computational load on each spectral element is calculated and finally, the one-dimensional array of spectral elements is partitioned, and elements are distributed to the processors so that the computational load is evenly distributed. Thus, if the particles are clustered in an area of the domain at the start of simulation, then processors receiving elements from that area will be assigned fewer elements compared to processors receiving elements outside that area. As the particles begin to spread out, the current element to processor assignment becomes sub-optimal degrading performance and requiring a fresh element to processor assignment. In contrast to Nek5000, CMT-nek can perform a reassignment.\eatme{ of elements to processors.}

Dynamic load balancing for scientific applications that consist of coupled data structures is a challenging problem \cite{choudhary1992software}. In this paper, we showcase the improvement in performance of CMT-nek (with moving particles and multiple coupled data structures) upon using dynamic load balancing, with  minimal overhead added due to the load balancing step itself. The application has a potential of scaling to millions of MPI ranks in the future and load balancing enhances its scalability. This is the type of scaling that is required for achieving high performance on next generation exascale machines. The load balancing techniques presented here may be applied to a broad class of mesh-based or otherwise multidomain numerical solvers of partial differential equations.

We developed three different algorithms for load balancing based on centralized, distributed and hybrid approaches, respectively, to distribute the computational load among processors. A pre-processing step uses an architecture independent re-ordering strategy to organize the load as a one-dimensional array \cite{hendrickson1993multidimensional} of spectral elements with particles while preserving spatial locality. The particles are mapped to the same processor that process the corresponding spectral elements containing these particles. The load balancing algorithm speeds up CMT-nek by a factor of up to $9.97$. On an Intel Broadwell platform, the overhead of the hybrid load balancing algorithm was only $1.82$ times the\eatme{computational }cost of each iteration for $65,520$ MPI ranks, whereas on a BG/Q platform, the overhead of the distributed load balancing algorithm was only $2.33$ times the computational cost of each iteration for $393,216$ MPI ranks. Given that load balancing is performed \eatme{only }every several hundred to several thousand iterations, the load balancing overhead is negligible.

We also developed an algorithm to automatically initiate a load balancing step as the application is running. This relieves the user from having to specify a fixed interval when load balancing should be triggered. We show that in one case this improves the overall time by about 9.4\% compared to a fixed load balancing approach where a load balancing interval is specified by the user.

The rest of the paper is organized as follows. Section~\ref{sec:relatedWork} presents the related load balancing literature and how our work differs from existing work. Section~\ref{sec:background} gives a brief background of CMT-nek. Section~\ref{sec:map} presents our load balancing strategies. Experimental results and conclusions are given in Sections~\ref{sec:experiments} and \ref{sec:conclusion}, respectively.

\vspace{-0.4cm}
\section{Related Work} \label{sec:relatedWork}
Dynamic load balancing has been researched extensively in various other applications which require high performance computing. For example, Lieber et al. \cite{lieber2017highly} decouple the cloud scheme from the static partitioning of the atmospheric model. Essentially, the cloud data are managed by a new, highly scalable framework, that supports dynamic load balancing. Due to the presence of tightly coupled data structures of fluid and particles, CMT-nek cannot use that strategy of creating a specialized framework for particles.

Menon et al. \cite{menon2012automated} present an adaptive load-balancing strategy where the application monitors and decides the right time to trigger a load-balancing scheme. While their work focuses specifically on automatically triggering load balancing, in this work, we developed a comprehensive load-balancing strategy for a real application, which includes an automatic load balancer. The method used in our automatic load balancer determines the slope of performance degradation at runtime and based on this information the load balancing period $I$ is adjusted. Thus, unlike~\cite{menon2012automated}, our load-balancing period changes adaptively at runtime and hence is sensitive to performance variations during simulation. Our method may also trigger load balancing earlier in $I$ if for some reason performance degradation is greater than the overhead to load balance.

Another body of work involving load balancing is plasma-based particle-in-cell (PIC) code \cite{surmin2015dynamic, ferraro1993dynamic, plimpton2003load}. Load balancing these codes presents the same challenges as load balancing CMT-nek, due to the presence of tightly coupled data structures. CMT-nek is more complex as it always models particle-particle interactions (four-way coupling), whereas the PIC models such interactions only in certain cases when a collision operator is implemented. Compared to the work in \cite{germaschewski2013plasma} that uses the Hilbert-Peano curve to partition a domain, we use the spectral bisection method. Plimpton et al. \cite{plimpton2003load} as well as Nakashima et al. \cite{nakashima2009ohhelp} perform a spatial decomposition as in CMT-nek; however, the assignment of grid cells to processors is static, whereas the number of particles is balanced out as part of their load-balancing algorithm. In CMT-nek, on the other hand, the processors owning the particles also own the grid cells (or spectral elements, as we call them) and mapping of grid cells to processors changes every time load balancing is done.

Pearce et al. \cite{pearce2014load} balance out particle-particle interactions instead of the number of particles, in keeping with the fact that the work done by a processor is proportional to the interactions computed. This scheme benefits applications where\eatme{in which }the density of interactions is nonuniform. CMT-nek does not only simulates particle-particle interactions, but incurs additional computational load due to the interactions between particles and fluid. Thus, the methods developed by Pearce et al. may not readily be applied to CMT-nek.

Bhatele et al. \cite{bhatele2009dynamic} present a load-balancing approach suitable for molecular dynamics applications that have a relatively uniform density of molecules. The variation in particles density in CMT-nek is much different from the situation considered in Bhatele et al.

The use of parallel k-d trees can improve the cost of particle-particle interactions using techniques described in \cite{al2000parallel}. When the number of cores in the platform is very large, the underlying many-to-many communication between cores can be further optimized by methods described in \cite{ranka1995many}. These will be considered in future. \eatme{our future work.}

\vspace{-0.4cm}
\section{Background\label{sec:background}\vspace{-0.1cm}}
\subsection{CMT-nek}
CMT-nek solves the three-dimensional Euler equations of fluid dynamics in strong conservation-law form, using the discontinuous Galerkin spectral element method \cite{HacklCMTCAF} integrated by the third-order total-variation-diminishing Runge-Kutta scheme of Gottlieb and Shu \cite{TVDRK}. Banerjee et al. \cite{Banerjee-7839682} describe these steps in more detail, both for CMT-nek and its proxy CMT-bone. We will now describe the support for particles in CMT-nek.

\vspace{-0.3cm}
\subsection{Particles in CMT-nek}
The particle algorithm follows the evolution of each particle in a series of time steps. For a given time step, there are essentially three phases:

\vspace{-0.3cm}
\paragraph{Interpolation phase} Utilizing barycentric Lagrange interpolation~\cite{berrut2004}, fluid properties are interpolated from the grid points to the location of a particle. This is done on an element-by-element basis, meaning that only the Gauss-Lobatto grid points within the spectral element in which the particle resides contribute to the fluid properties at the particle location.

\vspace{-0.3cm}
\paragraph{Equation solve phase} The force from the fluid on a particle is evaluated using the previously interpolated fluid properties at the particle location, along with time-varying and constant properties of each particle, such as its velocity and mass. These forces are then used in explicit updates of the particle velocity and position using the same third-order Runge-Kutta formulation for the time integration as is used by the DGSEM solver for the fluid phase in the Eulerian reference frame.

\vspace{-0.3cm}
\paragraph{Movement phase} After the position of each particle is updated, it is possible that it can spatially reside within a different spectral element than it was in previously. If this occurs and the core that holds the data of the previous spectral is different than the core that holds the data of the new spectral element, the particle data is transferred to the new core.

\vspace{-0.4cm}
\subsection{Parallelization\label{sec:strategy} in CMT-nek}
\begin{figure}
\vspace{-0.5cm}
\includegraphics[height=1.7in, clip=true, trim=2cm 20cm 2cm 1.5cm]{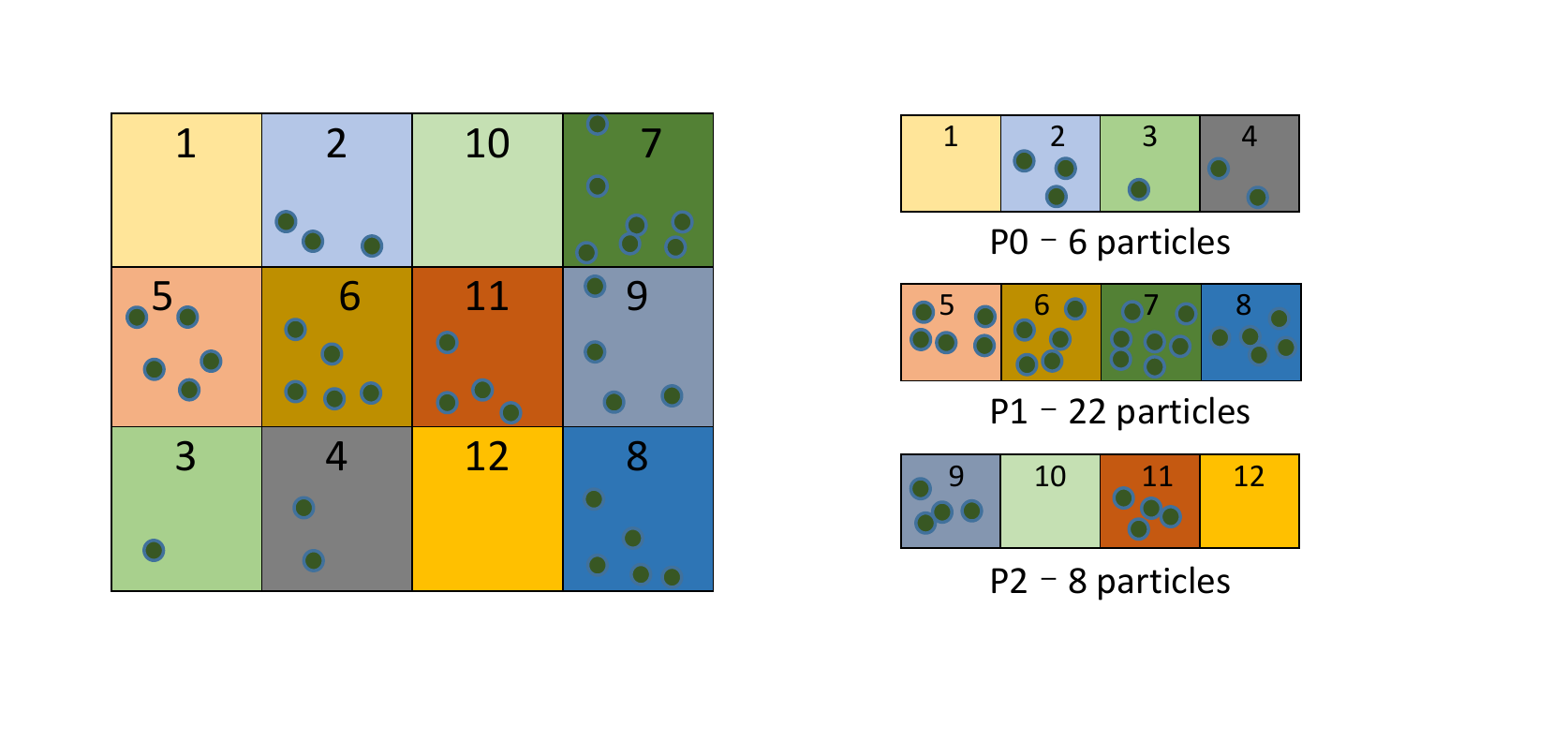}
\vspace{-1.6cm}
\caption{An example of using spectral bisection method to map a 2-D domain to a 1-D array and uniformly assign its elements to each processor. }
\vspace{-0.4cm}
\label{fig:mapel}
\end{figure}

To make an application scalable, the domain is partitioned, and each processor gets a section of the domain~\cite{Ou:1997:PRA:264305.264312, Ou1996}. The mapping process of spectral elements to processors may be decomposed into three main stages:
\begin{enumerate}
\item Transforming the 3-D distribution of elements in the fluid domain to an architecture independent 1-D array of elements while maintaining spatial locality~\cite{Ou:1995:ALT:224538.224573}
\item Partitioning the 1-D element array
\item Mapping the element partitions to the processors
\end{enumerate}
CMT-nek uses a recursive spectral bisection method~\cite{hendrickson1993multidimensional,Tufo:2001:FPD:372836.372845} to create the 1-D element array at first\eatme{. The elements in the 1-D element array  }, which are then partitioned such
that each processor gets an equal number of elements.
The particles contained in an element are then assigned to the processor to which the element is mapped. Figure~\ref{fig:mapel} shows an example where a 1-D element array is created from elements in a 2-D fluid domain and partitioned uniformly by CMT-nek, irrespective of the number of particles in an element. This partitioning strategy causes load balancing issues, especially in the case of particle-laden explosively driven flows, where billions of particles might be initially contained in a small region of the overall domain.

\begin{algorithm}[!t]
\caption{Compute load on an element}
\label{computeload}
\begin{algorithmic}[1]
    \Ensure An array of computational loads of elements, indexed by local element index
    \Function {compute\_element\_load}{}
        \For{each local element e}
            \State $elementLoad$[e] = particle load + fluid load
           \State \hspace{1cm} = number of particles in element + $fluid\_load$ \Comment where $fluid\_load$ = average time to process an element / average time to process a particle
        \EndFor
    \EndFunction
\end{algorithmic}
\end{algorithm}
\vspace{-0.2cm}

\setlength{\textfloatsep}{9pt}
\begin{algorithm}[htb]
\caption{Centralized load balancing algorithm}
\label{centralizedlb}
\begin{algorithmic}[1]
    \Ensure A load balanced element to processor assignment
    \Function {recompute\_partitions}{} 
    \For {each processor $P_i$}
        \State call \Call{compute\_element\_load}{}
    	\State send array {$elementLoad$} to Processor $P_0$
    \EndFor
    \State $P_0$ receives {$elementLoad$} from processors 
    \State $P_0$ sorts {$elementLoad$} based on global element index
    \State $P_0$ computes prefix sum of {$elementLoad$}
    \State $P_0$ stores the prefix sum in array {$prefix\_sum$} 
    \State  call \Call {partition\_load}{$prefix\_sum$, $length$} to decide where to chop the {$prefix\_sum$} array and assign element$\rightarrow$processor map   
    \State $P_0$ broadcasts the new element$\rightarrow$processor map
    \State \Return element$\rightarrow$processor map
    \EndFunction  
\end{algorithmic}
\end{algorithm}

\section{Load Balancing CMT-nek}\label{sec:map}
In this section we describe our load balancing strategy. 
Section~\ref{sec:computeload} shows how the computational load on an element is determined for the purposes of load balancing. Section~\ref{sec:repartition} presents a centralized, a distributed and a hybrid version of the domain repartitioning algorithm. Section~\ref{sec:example} presents an example that further clarifies the steps in these algorithms. Section~\ref{sec:transfer} describes the logistics of data sharing between processors, and finally, Section~\ref{sec:trigger} describes two algorithms for triggering load balancing at runtime. 

\vspace{-0.4cm}
\subsection{Determining computational load on a spectral element\label{sec:computeload}}
Computational load on an element is quantified as the number of particles present inside the element plus a baseline load for fluid\eatme{(either liquid or gas)} computation. The latter is considered to be about the same for any element since fluid computations involve solving fluid properties at each grid point in an element and all hexahedral spectral elements have the same number of grid points. In fact, the load on each element due to fluid computations is $O(N^4)$, where $N$ is the number of grid points along one direction~\cite{gapaper}. To ensure that particle load does not dominate fluid load and vice versa, we represent fluid load as a constant defined as the ratio of the average time it takes to process a single element to the average time it takes to process a single particle by running the application prior to enabling load balancing.
This constant is dependent on the platform and also on problem parameters such as grid size. Algorithm~\ref{computeload} describes the computation of relative load on an element. Once the computational load for each element is determined, the next step is to repartition the 1-D array of spectral elements as shown in Figure~\ref{fig:mapel}, which is described in next section.

\begin{algorithm}[htb]
\caption{Partition algorithm called by centralized load balancing algorithm}
\label{alg:ldblce_cen}
\begin{algorithmic}[1]    
    \Function {partition\_load}{$prefix\_sum, length$}
    \Comment this function is called in Algorithm \ref{centralizedlb}, $length$ is the number of entries in the $prefix\_sum$ array
    \State $np$ = number of processors \Comment create $np$ partitions of element array as follows
    \State $lp$ = 0 \Comment position of last partitioning
    \For{i = 1, $np$ - 1}
        \State $threshold = i\times prefix\_sum(length)/ np$
        \State iterate over $prefix\_sum$ and determine the
        \State \hspace{0.4cm}position $p$ of the first sum that exceeds $threshold$
        \State consider $prefix\_sum$ entries at $p$ and $p-1$
        \State $d_1$ = distance($prefix\_sum(p-1)$, $threshold$)
        \State $d_2$ = distance($prefix\_sum(p)$, $threshold$)
        \If{$d_1$ $<$ $d_2$ } 
        \State include $p-lp-1$ entries in current partition
        \State $lp$ = $p-1$
        \Else 
        \State include $p-lp$ entries in current partition
        \State $lp$ = $p$
        \EndIf
        \State map elements in current partition to processor $P_{i-1}$
    \EndFor
    \State map elements in the last partition to the last processor
    \State \Return element $\rightarrow$ processor map
    \EndFunction
\end{algorithmic}
\end{algorithm}

\vspace{-0.4cm}
\subsection{Repartitioning strategies\label{sec:repartition}}
We developed a centralized, a distributed and a hybrid algorithm for repartitioning the 1-D array.

\vspace{-0.2cm}
\subsubsection{Centralized}
The main theme of the centralized algorithm is that all the processors send their total computational load to processor $P_0$. Processor $P_0$ then computes the prefix sum of the load on elements ordered according to their global IDs, partitions the prefix sum array, uses the partitioning to create a new element$\rightarrow$processor map $M'$, and finally, broadcasts $M'$ to all processors. This scheme is presented in Algorithm~\ref{centralizedlb} and the steps are traced using an example.

\vspace{-0.2cm}
\subsubsection{Distributed}
The centralized load balancing has processor $P_0$ in the critical path that determines how fast the load balancing may complete. In the distributed version, we remove this bottleneck and let each processor collaborate to have a local copy of the prefix sum of the load. After that, each processor calculates a local element$\rightarrow$processor map. The processors share their local maps which each processor composes to form a global element$\rightarrow$processor map. As a last step, each processor adjusts the mapping to guarantee that the number of elements assigned to a processor does not exceed a maximum bound defined by the user by setting a variable ``lelt". The distributed load-balancing scheme is presented in Algorithm~\ref{distributedlb} and the steps are traced using an example.

\vspace{-0.3cm}
\subsubsection{Hybrid}
The hybrid load-balancing algorithm is a combination of the centralized and distributed algorithms. First, the processors collaboratively create the local copy of the prefix sum and each processor calculates a local element$\rightarrow$processor map. Then, each processor sends the local map to processor $P_0$. $P_0$ then aggregates the data to create the global map. $P_0$ also adjusts the mapping to guarantee that the number of elements assigned to a processor does not exceed ``lelt". Finally, processor $P_0$ broadcasts the element$\rightarrow$processor map, $M'$, to all processors. In order to save space, the hybrid algorithm is omitted here. However, it is basically Lines 1-7 in Algorithm \ref{distributedlb}, Lines 1-20 in Algorithm \ref{alg:ldblce_dist}, sending element$\rightarrow$processor map to processor $P_0$, Line 22 in Algorithm \ref{alg:ldblce_dist} executed on processor $P_0$, and finally, broadcasting the new element$\rightarrow$processor map to all processors.

\begin{algorithm}[htb]
\caption{Distributed load balancing algorithm}
\label{distributedlb}
\begin{algorithmic}[1]
    \Ensure A load balanced element to processor assignment
    \For {each processor $P_i$ in parallel}
        \State call \Call{compute\_element\_load}{}
    	\State compute {$prefix\_sum$} for element load array of $P_i$
        \State $loadsum_i$ = sum of total load on $P_0$, $\cdots$, $P_{i-2}$, $P_{i-1}$
        \For {each entry in {$prefix\_sum$} array}
            \State add $loadsum_i$ to that entry
        \EndFor
        \State call \Call {partition\_load\_distributed}{$prefix\_sum$,$nelgt$,$lelt$}
    	\EndFor	  
\end{algorithmic}
\end{algorithm}
\vspace{-0.4cm}

\begin{algorithm}[htb]
\caption{Partition algorithm called by distributed load balancing algorithm}
\label{alg:ldblce_dist}
\begin{algorithmic}[1]     
    \Require Prefix sum of element load array, {$prefix\_sum$}
    \Require Total number of elements in the application, $nelgt$
    \Require Maximum number of elements allowed on a processor, $lelt$
    \Ensure Element to processor mapping
     \Function {partition\_load\_distributed}{$prefix\_sum$, $nelgt$, $lelt$}
         \Comment This function is called in Algorithm \ref{distributedlb} by processor $P_i$
         \State $np$ = number of processors
         \State $total\_load$ = total number of particles + $fluid\_load \times nelgt$ 
         \State $loadavg$ = $total\_load / np$
         \Comment Each processor should ideally have $loadavg$ load; first round of processor assignment:
         \State $e$ = local element index
         \For {$e$ = 0 to $last\_e$ } \Comment all elements in $P_i$
         \State $processor[e]$ = ($prefix\_sum[e]$ - 1) / $loadavg$
         \EndFor\\
         \Comment Second round of processor assignment to ensure the number of elements assigned to a processor does not exceed limit $lelt$
         \If{$P_i$ is neither the first nor the last processor}
             \State Send $processor[last\_e]$ to $P_{i+1}$
             \State Recv $processor[last\_e]$ of $P_{i-1}$ 
             \State Store received value in $recvAssign$
         \ElsIf{ current processor is the first processor}
             \State Send $processor[last\_e]$ to $P_{i+1}$
         \ElsIf {current processor is the last processor}
         \State Recv $processor[last\_e]$ of $P_{i-1}$
         \State Store received value in $recvAssign$
         \EndIf
         \State Using $recvAssign$ when present, and $processor[e]$ mapping for all local elements, $P_i$ creates a mapping of processor index and element left boundary represented by the global element index of first element assigned to the processor.
         \State The processors call $MPI\_ALLGATHERV$ to gather this information and at the end of this step all processors have a mapping of processor index and global element index of the first element.
         \State $P_i$ checks if the number of elements assigned to processor $P_k, k=0,\cdots, np-1$, is bounded by $lelt$. If number of elements assigned to processor $P_k$ is greater than $lelt$, the right element boundary for $P_k$ is moved to left until the number of elements in $P_k$ is $lelt$. The number of elements in $P_{k+1}$ increases and the bound check is continued for processor $P_{k+1}$, $P_{k+2}$ and so on.
     \EndFunction
\end{algorithmic}
\end{algorithm}

\begin{figure}[htb]
\vspace{-0.5cm}
\includegraphics[height=2.6in, clip=true, trim=2.8cm 15.5cm 2cm 2cm]{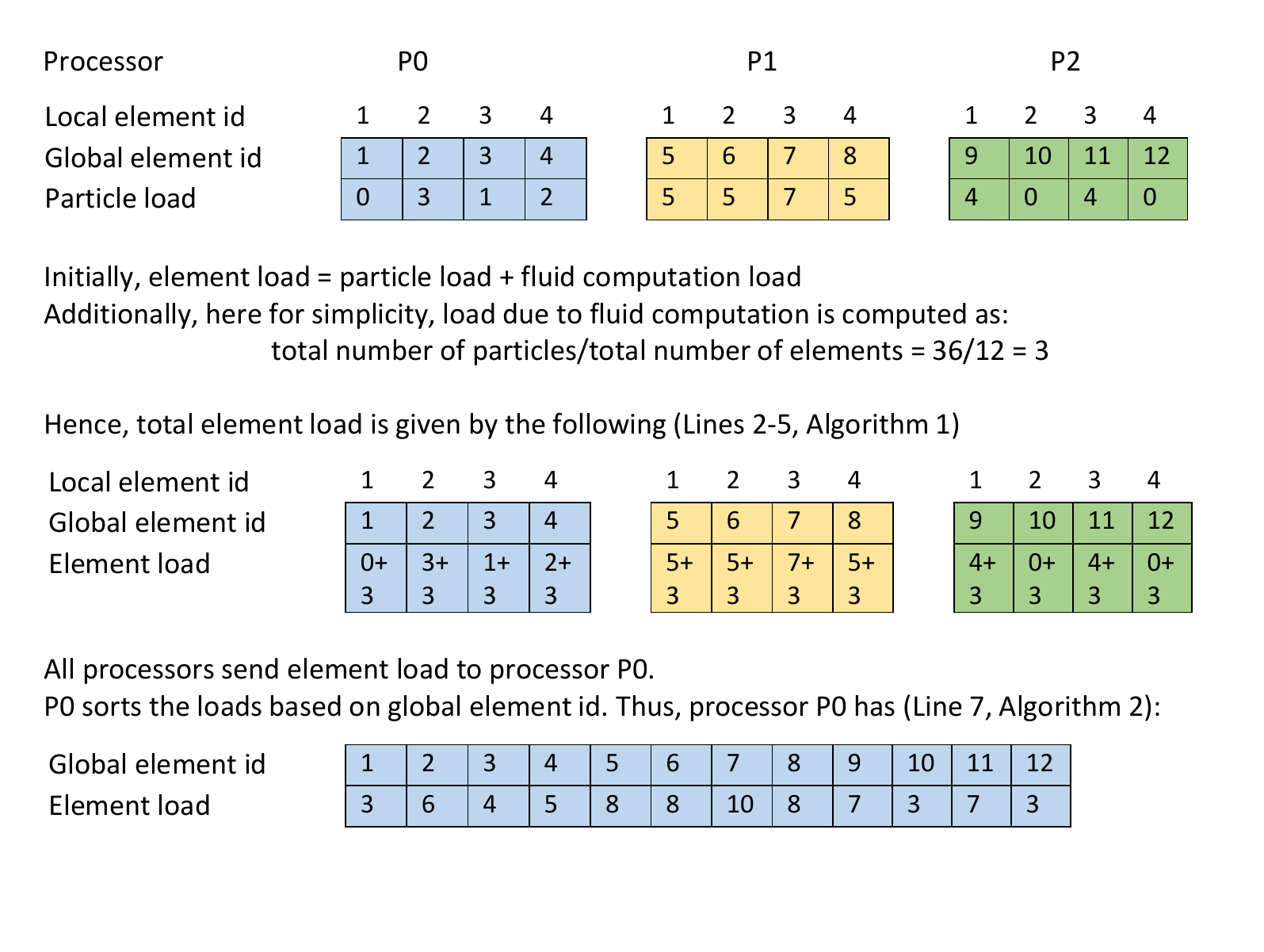}
\vspace{-1.4cm}
\caption{Centralized LB Step 1: Calculate element load array and send it to processor $P_0$}
\label{fig:cenlbstep1-3}
\vspace{-0.4cm}
\end{figure}

\subsection{An example illustrating the repartitioning strategies\label{sec:example}}
Suppose the fluid domain has been partitioned into $12$ elements, converted into 1-D array using the recursive spectral bisection method, and that there are $36$ particles placed in these elements, as shown in Figure~\ref{fig:mapel}. Further suppose that there are $3$ processors, $P_0$, $P_1$ and $P_2$. Initially, CMT-nek  partitions the element array uniformly, so each processor gets $4$ elements. This setup and the initial element$\rightarrow$processor assignment are shown in Figure~\ref{fig:cenlbstep1-3}. Using this setup, the element load is calculated using Algorithm~\ref{computeload}, where fluid load is computed as $3$.

\textbf{\textit{Steps in centralized load balancing.}}
In a centralized version, all processors send their arrays of element load to a central processor $P_0$ as shown in Figure~\ref{fig:cenlbstep1-3}. We use the underlying crystal router mechanism in Nek5000 for this communication. Thus, by the end of this step, the processor $P_0$ has gathered the load for all elements and have sorted them based on the global element ordering.

\begin{figure}[htb]
\vspace{-0.5cm}
 \includegraphics[height=4.1in, clip=true, trim=2.95cm 10.5cm 2cm 1.7cm]{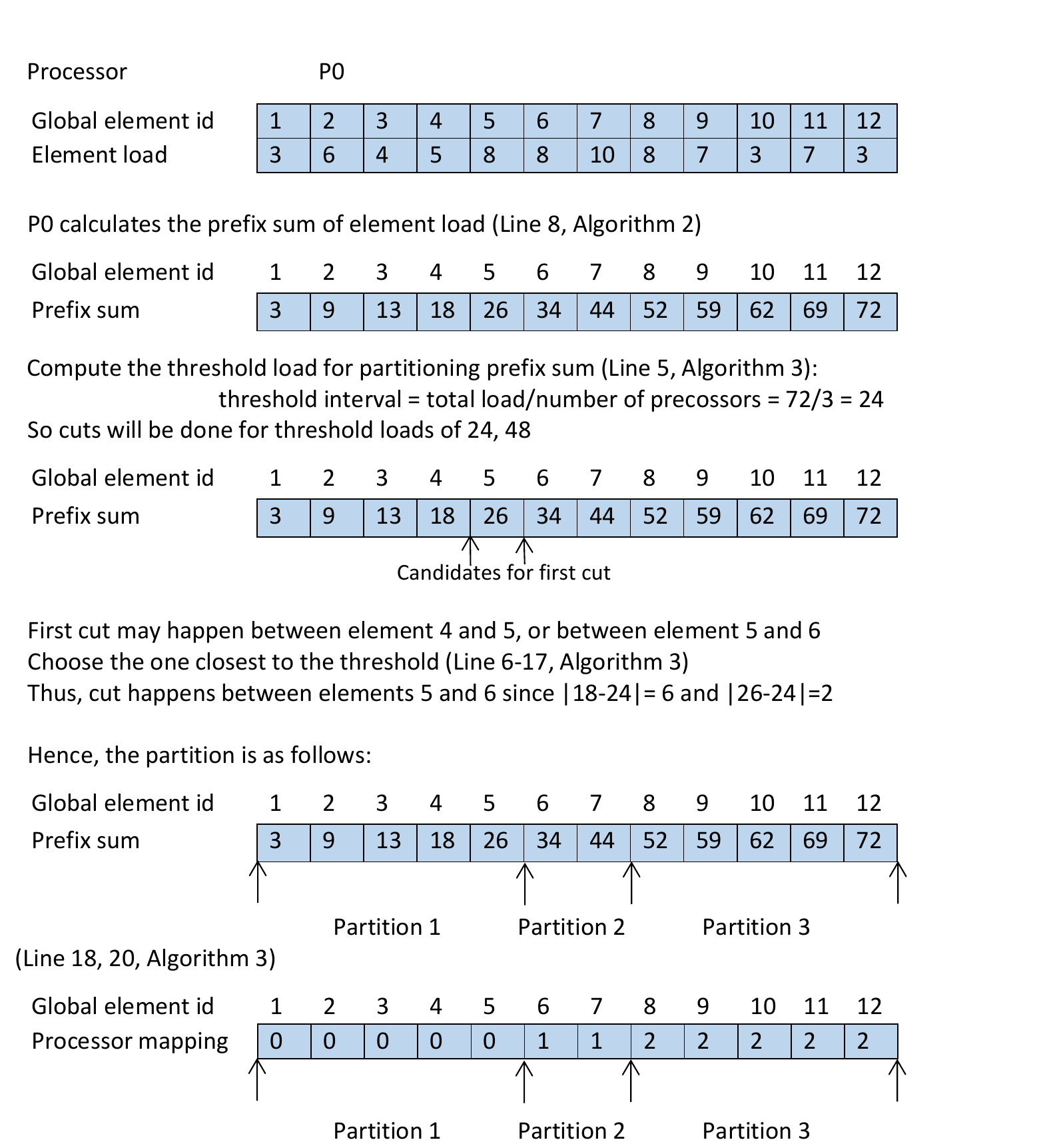}
\vspace{-0.7cm}
\caption{Centralized LB Step 2: Processor $P_0$ determines new element$\rightarrow$processor mapping array based on the prefix sum of element load}
\vspace{-0.4cm}
\label{fig:cenlbstep4}
\end{figure}

\begin{figure}[htb]
\includegraphics[height=0.68in, clip=true, trim=3cm 24cm 2cm 2.6cm]{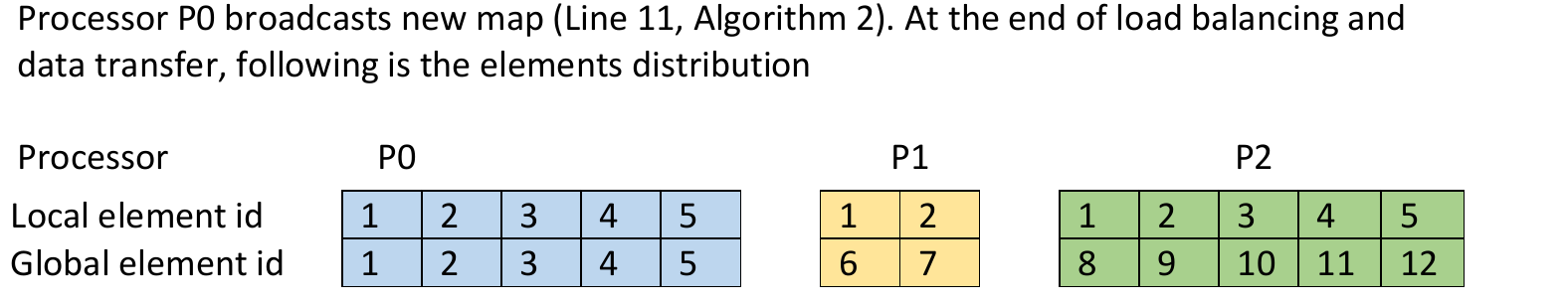}
\vspace{-0.6cm}
\caption{Centralized LB Step 3: Every processor gets the new partition after load balancing}
\label{fig:cenlbstep5}
\vspace{-0.1cm}
\end{figure}

\begin{figure}[htb]
\vspace{-0.3cm}
\includegraphics[height=2.3in, clip=true, trim=3cm 16.5cm 2cm 2.6cm]{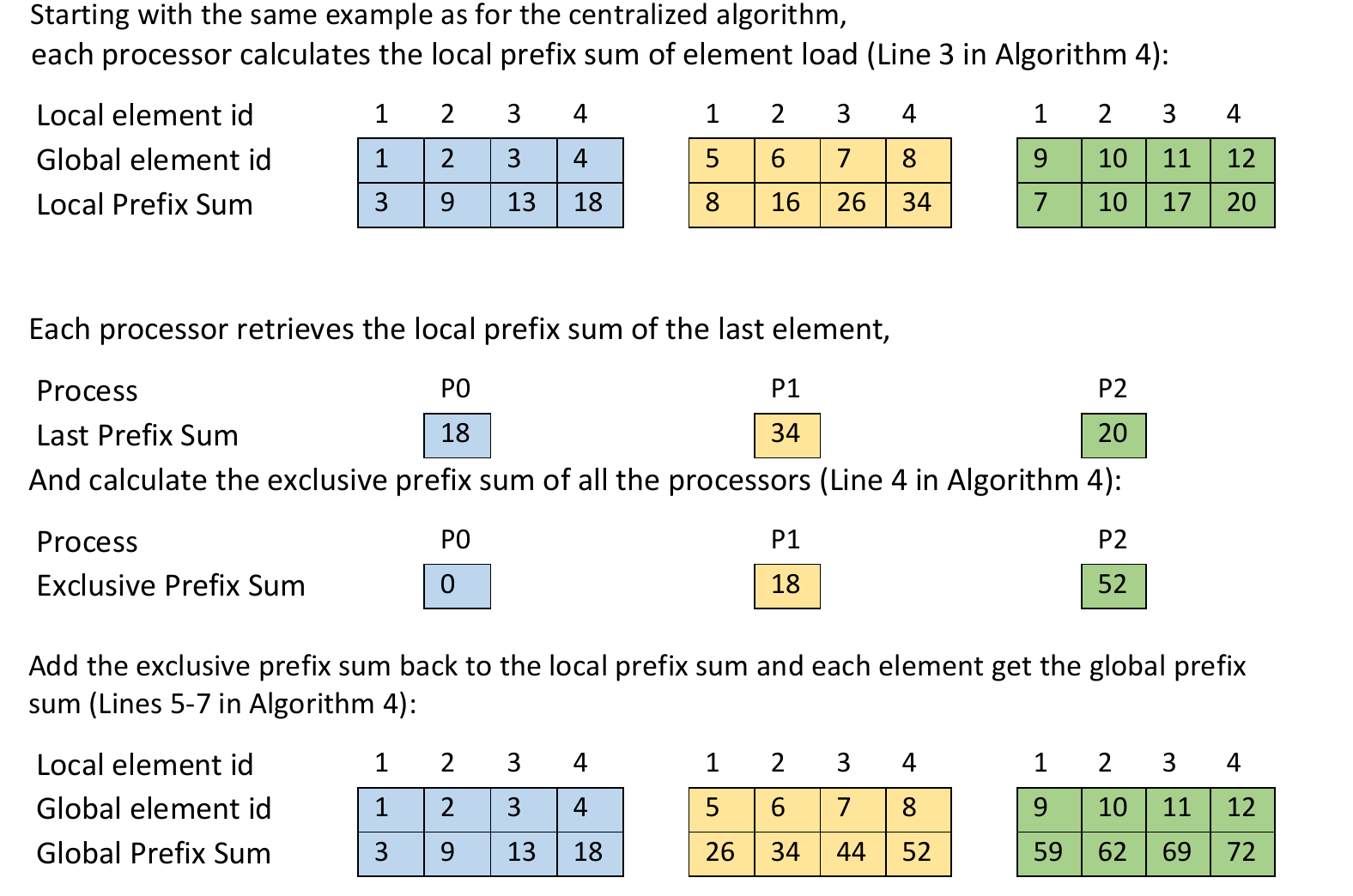}
\vspace{-0.7cm}
\caption{Distributed LB Step 1: Calculate element load and the global prefix sum}
\label{fig:distlb-step1}
\vspace{-0.4cm}
\end{figure}

Figure~\ref{fig:cenlbstep4} shows the second step in the centralized algorithm, where $P_0$ processes the element load information received. The element load array stores the total computational load on elements indexed by the global element index. Following that, processor $P_0$ computes its prefix sum, as well as a  threshold load computed as the total load divided by the number of processors. The threshold load helps us to distribute the load as evenly as possible on the processors.
In our example, the threshold load is $72/3 = 24$. 

After a threshold value is calculated, processor $P_0$ iterates over the element load array and checks the point where the threshold is reached or just exceeded. When that point is reached, two preceding values in the element load array are evaluated to check where the partition should be placed. In our example, the prefix sum values for elements $4$ and $5$ (sum values of $18$ and $26$) are evaluated against the threshold to determine the first position of partitioning. Since $24-18=6$, whereas $26-24=2$, element $5$ is nearer to threshold $24$, and therefore, the position of the first cut is in between the fifth and sixth elements. In the meantime, ``lelt" is also considered. If the maximum number of elements in a processor has been reached while the threshold position hasn't, the partition is still placed here.

Since there are three processors, two cuts will be needed to create three partitions. The $n^{th}$ cut to the load array will happen at a point where the load is close to $threshold \times n$. Thus, the threshold for the second cut is $24 \times 2 = 48$, and the second cut is made between the seventh and eighth elements.

\begin{figure}[htb]
\includegraphics[height=3.8in, clip=true, trim=3cm 8.9cm 2cm 2.6cm]{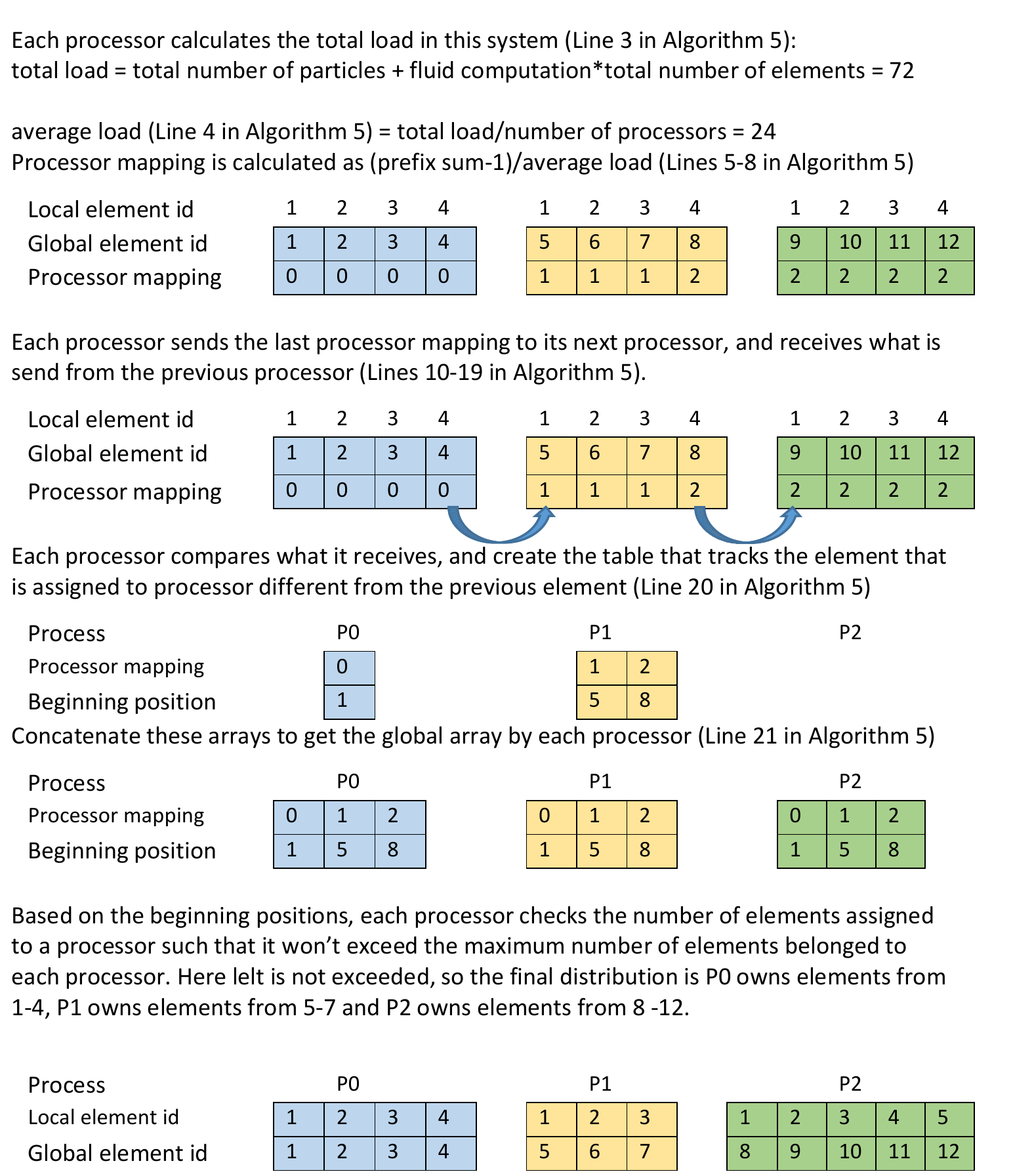}
\vspace{-0.7cm}
\caption{Distributed LB Step 2: Each processor comes up with the new element$\rightarrow$processor mapping array}
\vspace{-0.3cm}
\label{fig:distlb-step2}
\end{figure}

Once the partitions are made, processor $P_0$ creates the element $\rightarrow$ processor map. As shown in Figure~\ref{fig:cenlbstep4}, the first five elements in the global 1-D array of elements are assigned to processor $P_0$, the next two elements are assigned to processor $P_1$ and the last five elements are assigned to processor $P_2$. This processor mapping information, is with processor $P_0$ only at this point. The mapping array is then distributed by $P_0$ to all the remaining processors. This step concludes the centralized algorithm to remap elements to processors as shown in Figure~\ref{fig:cenlbstep5}. 

\textbf{\textit{Steps in distributed load balancing.}}
In the distributed version, each processor computes the loads of elements assigned to it, followed by a local prefix sum of the loads as shown in Figure \ref{fig:distlb-step1}. To convert a local prefix sum to a global prefix sum, each processor $P_i$ needs to know the prefix sum of load residing on processor $P_0$ through processor $P_{i-1}$, which is also known as exclusive prefix sum and implemented using $MPI\_EXSCAN$. By adding this exclusive prefix sum to each entry of the local element load array, the global prefix sum is obtained. This process is illustrated in Figure \ref{fig:distlb-step1}.

After obtaining the global prefix sum, each processor computes element$\rightarrow$processor mapping as shown in Figure \ref{fig:distlb-step2}. First, the total and average \eatme{\textcolor{red}{processor} global element }loads are calculated, which are 72 and 72/3 = 24, respectively, in our example. Given that the processors are homogeneous, the element$\rightarrow$processor mapping may be obtained by simply dividing each global prefix sum (see Figure~\ref{fig:distlb-step1}), less 1, by the average load. At this point, each processor has a portion of the new element$\rightarrow$processor map, shown as ``Processor mapping" in Figure~\ref{fig:distlb-step2}. Before making this new mapping the final one,  we need to guarantee that the number of elements assigned to a processor does not exceed the maximum number defined by the user, that is ``lelt". A naive way to check it is using an all-to-all communication to send and construct the entire global element$\rightarrow$processor map at each processor and adjusting the number of elements such that no processor is assigned more than ``lelt" elements.

Instead of sending the whole local element$\rightarrow$processor map using the all-to-all communication as described above, the following happens in our implementation: 1) each processor sends the processor mapping of the last element to the next processor, and receives the data sent by the previous processor; and 2) each processor compares data received from the previous processor with the processor mapping for the first local element, to check if the processor mapping is the same or not. For example in Figure~\ref{fig:distlb-step2}, the data received by $P_2$ is $2$, and it is the same as the processor mapping assigned to its first local element. Similarly, the data received by $P_1$ is $0$, which is different from the processor number assigned to the first local element in $P_1$. This is significant because this information is used by each processor to identify the global element index of the first element assigned to any processor. For example, $P_2$ can tell that the element with global index $9$ is not the first element assigned to $P_2$. Similarly, $P_1$ can tell that the global element index of the first elements assigned to $P_1$ and $P_2$ are $5$ and $8$, respectively. Then each processor collectively has a list of the global element indices of the first elements assigned to processors. Only this information is then shared among the processors using all-to-all communication, and finally, all processors have a consolidated list of global element indices of the first elements assigned to processors. Using this consolidated list, the processors then check the maximum bounds on number of assigned elements (``lelt"), and fix it if violated. This step concludes the distributed algorithm to remap elements to processors as shown in Figure \ref{fig:distlb-step2}.

\textbf{\textit{Steps in hybrid load balancing.}} In order to save space and avoid repetition, the examples for the hybrid load balancing is omitted here. It is the same as the distributed load-balancing examples in Figure \ref{fig:distlb-step1} and the first 3 steps in Figure \ref{fig:distlb-step2}. For the last but one step in Figure \ref{fig:distlb-step2}, only processor $P_0$ receives the processor mapping and adjusts the position to make the number of elements in a processor within ``lelt". Then, processor $P_0$ broadcasts the new element$\rightarrow$processor map to all other processors.

Though in this example we started with uniform partitioning and defined a repartitioning strategy, the same process may be repeated as needed during the course of simulation whenever load imbalances arise.

The readers would note that the element$\rightarrow$processor mapping arrays obtained using the centralized and distributed algorithms are different. That is because for the centralized algorithm, processor $P_0$ has global information of each element's load. While for the distributed algorithm, each processor only has local information. Thus, the centralized algorithm can make better decisions with regard to the element$\rightarrow$processor mapping, compared to the distributed algorithm. However, there is no performance bottleneck caused by a single node in the distributed algorithm as in the centralized algorithm. Thus, there is a trade-off between better decisions and performance. The benefits of a hybrid load-balancing algorithm would be explained in Section \ref{subsec:expQuartz}.

\textbf{\textit{Transfer of elements and particles.}}
The new element $\rightarrow$ processor map is used by the processors to transfer elements and particles appropriately. For example, based on the new map obtained by the centralized algorithm in our example, processor $P_1$ would send it's first and last elements to processors $P_0$ and $P_2$, respectively. The particles contained in the domain of each transferred element are transferred to the same respective destination processors.

So far we have described computational load balancing of elements and particles on processors. It is also important to consider imbalances generated in the communication load. CMT-nek follows a discontinuous Galerkin scheme~\cite{Banerjee-7839682,HacklCMTCAF} and hence is light on communication. Communication is required for transferring shared faces of elements residing on different processors. The original CMT-nek distributes elements uniformly among the processors. Hence, the number of faces to be shared is bounded implicitly. On the other hand, load-balanced CMT-nek distributes elements nonuniformly, which in the worst case, may result in an increase in the communication overhead. The communication overhead for a processor is upper bounded in the load-balanced code by specifying a maximum limit on how many elements may be assigned to the processor (``lelt"). Having such a limit also ensures that all elements and data fit into processor memory.

\vspace{-0.2cm}
\subsection{Distributing Elements and Particles\label{sec:transfer}}
In this section we will describe the details of how data is transferred between processors. There are a large number of data structures in CMT-nek that contain element and particle information. Some of these data structures store static data while others store dynamic data. Static data include information such as the $x$, $y$ and $z$ coordinates of each element, curvature on the curved faces, and the boundary conditions. Dynamic data includes fluid and particle properties that change during simulation.
We follow two primary strategies for information transfer. 

\vspace{-0.2cm}
\subsubsection{Transferring data} Arrays storing the conserved variables such as mass, energy, and the three components of momentum are transferred. The transfer process consists of packing the arrays to be transferred, transferring the packed array, and finally unpacking the arrays received. We use the underlying crystal router in Nek5000 for transferring the packed data.

\vspace{-0.2cm}
\subsubsection{Reinitializing data} The data structures which store static data are reinitialized. For this step, we initiate calls to existing Nek5000 data initialization routines. This needed careful analysis since some routines store the status of calls using static variables local to the scope of the routines. We analyzed and updated such routines employing static variables to reset the state of those variables when load balancing is done.

After load balancing is complete, a processor will start the next time~step, which involves the computation of field variables for all elements, including the new ones that were received. For all our tests we have diligently verified that the results of simulation has the same accuracy as the original CMT-nek.

\vspace{-0.3cm}
\subsection{Triggering a load-balancing step\label{sec:trigger}}
As particles move during simulation, the original element $\rightarrow$ processor mapping becomes suboptimal since the particle-heavy elements with large computational load start getting lighter on particles as the particles move apart. This motivates the need for a dynamic load-balancing scheme, where load balancing may be triggered during an ongoing simulation process. There are two main strategies for triggering load balancing during simulation in CMT-nek. The first is performing load balancing at specific intervals where the intervals are specified by the user. The second strategy is adaptive load balancing, where no input is necessary from the user. 

\vspace{-0.2cm}
\subsubsection{Fixed step load balancing} This type of load balancing requires user input. For example, a load-balancing step may be triggered after every $k$ time steps, where $k$ is specified by the user. To set a reasonable value for $k$, the user should be aware of the simulation details, such as the problem size, particle speed, duration of a time step, and so on. The user may also run the simulation for a certain time without load balancing to estimate $k$. 

\begin{algorithm}[!t]
\caption{Automatic load balancing}
\label{fig:adaptivelb}
\begin{algorithmic}[1]
    \Function {Adaptive\_load\_balance}{}
    \Comment is called after solver step in the simulation.
    \State $threshold = 0.05;$ $  $  $degradation$ = 0.0
    \State $r\_step = 0$ \Comment Time step in which load balance happens
    \State $lb\_time$ = time taken by load balancing algorithm
    \State $eval\_interval$ = 100 \Comment Evaluate performance for these many steps. An evaluation phase $P$ starts one step after each time load balance happens, $P$ consists of $eval\_interval$ steps 
    \State $t_1$ = average time per time-step in $P$
    \State $c_1$ = middle step in $P$ \Comment $c_1$ = 50, here, initially
    \State $cts$ = current-time-step
    \State $t_2$ = median of time per time-step among $[$$cts-2$, $cts$$]$
    \State $lb\_once$ = false \Comment set to true after first call to load balance
    \If{$cts \in P$}
    \State Update $t_1$, $c_1$; $ $ continue;
    \ElsIf{$lb\_once$ == false}
            \If{$(t_2-t_1)/t_1 > threshold$}
    
            \State Perform load balance; calculate $lb\_time$; 
            
            \State $reinit\_itv = cts - c_1;$ $r\_step = cts$; $lb\_once$ = true
            \EndIf
    \ElsIf{$cts - r\_step == 1$}
            	\State $rebal=sqrt(2*reinit\_itv*lb\_time/(t_2-t_1))$ \Comment $rebal$ is the number of steps after which the next load balance would theoretically happen
               \Else
               \State $degradation = degradation + (t_2-t_1)$
               \If{$cts - r\_step \ge rebal$ $||$ $degradation > lb\_time$}
               \State Perform load balance; calculate $lb\_time$
               \State $reinit\_itv = cts - r\_step$
               \State $r\_step = cts;$ $degradation = 0.0$
               \EndIf
            \EndIf
    \EndFunction
\end{algorithmic}
\end{algorithm}

\vspace{-0.3cm}
\subsubsection{Adaptive load balancing} This type of load balancing is performed automatically by the program, with no input from the user. The details of the adaptive load-balance algorithm are presented in Algorithm~\ref{fig:adaptivelb}. Before the adaptive load balancing strategy is applied, there is one compulsory load-balancing step which happens at the beginning after particles are initialized and placed and before the start of simulation. After that, in Algorithm~\ref{fig:adaptivelb}, a load-balancing step happens whenever performance degrades by a certain threshold (Line 14-17 in Algorithm \ref{fig:adaptivelb}). For subsequent load-balancing steps, the application captures the cost of load balancing (i.e., $lb\_time$ in Algorithm \ref{fig:adaptivelb}), the slope of performance degradation (i.e., $(t2-t1)/reinit\_itv$ in Algorithm \ref{fig:adaptivelb}). According to \cite{menon2012automated}, we can get the theoretical load-balancing interval (Line 19 in Algorithm \ref{fig:adaptivelb}). When this interval is reached, the load-balancing algorithm is called. However, since the slope may be changing, we add another criteria that is when the cost of load balancing is covered by the cost caused by the increasing time-per-time-step, the load balancing is called. After that, each time a load balancing is performed, the cost of load balancing and the slope are updated and the new theoretical load-balancing interval is calculated. Thus, load balancing is called when either one of the two criteria is satisfied.

Note that, the value of threshold determines the time step when load balancing happens for the first time after simulation begins. The subsequent load balancing time steps is determined by calculating $rebal$. Our experiments show that a threshold between 0.03 to 0.2 works equally well with a standard deviation of 0.016 for average time per time step.

\vspace{-0.3cm}
\section{Experimental Results\label{sec:experiments}}
We begin this section by describing the test case and the platforms used for this study. Then we determine the cost of load balancing and how it scales. Finally, we discuss the improvement in performance of CMT-nek obtained using load balancing.

\vspace{-0.2cm}
\subsection{Problem description}\label{sec:testcase}
In order to test the load-balancing algorithm, we used a test case that has been devised to mimic some of the key features of particle-laden, explosively driven flows that the load-balancing algorithm proposes to overcome. The test deals with expansion fans in one dimension which are simple compressible flows.\eatme{that retain some fundamental traits distinguishing them from incompressible flows. Figure~\ref{fig:expfan} shows the geometry of an expansion fan.}The problem domain is a rectangular prism that extends from $0$ to $0.0802$ in the $y$ and $z$ directions and from $-2.208$ to $6.0$ in the $x$ direction. Note that the units in this case are non-dimensional. The particles are assigned between $-1.0$ 
and $-0.5$ in $x$ direction, where the difference between the left ($x=-1.0$) and right ($x=-0.5$) boundaries determines the initial volume fraction of particles. The left boundary is often adjusted to obtain a different initial volume fraction.

\vspace{-0.3cm}
\subsection{Platform\label{sec:platforms}}
\subsubsection{Quartz} Quartz is an Intel Xeon platform that is located in Lawrence Livermore National Laboratory. Quartz has a total of 2,688 nodes, each having 36 cores. Each node is a dual socket Intel 18-core Xeon E5-2695 v4 processor (code name: Broadwell)  operating at 2.1 GHz clock speed. 
The memory on each node is 128 GB. Quartz uses Omni-Path switch to interconnect with the parallel file system. 

\vspace{-0.2cm}
\subsubsection{Vulcan}Vulcan is an IBM BG/Q platform which is located in the Lawrence Livermore National Laboratory. Vulcan has 24,576 nodes with 16 cores per node, for a total of 393,216 cores. Each core is an IBM PowerPC A2 processor operating at 1.6 GHz clock speed. 
The memory on each node is 16GB. Vulcan uses a 5-D torus network to interconnect with the parallel file system.

\begin{figure}
\centering
\includegraphics[width=0.9\columnwidth]{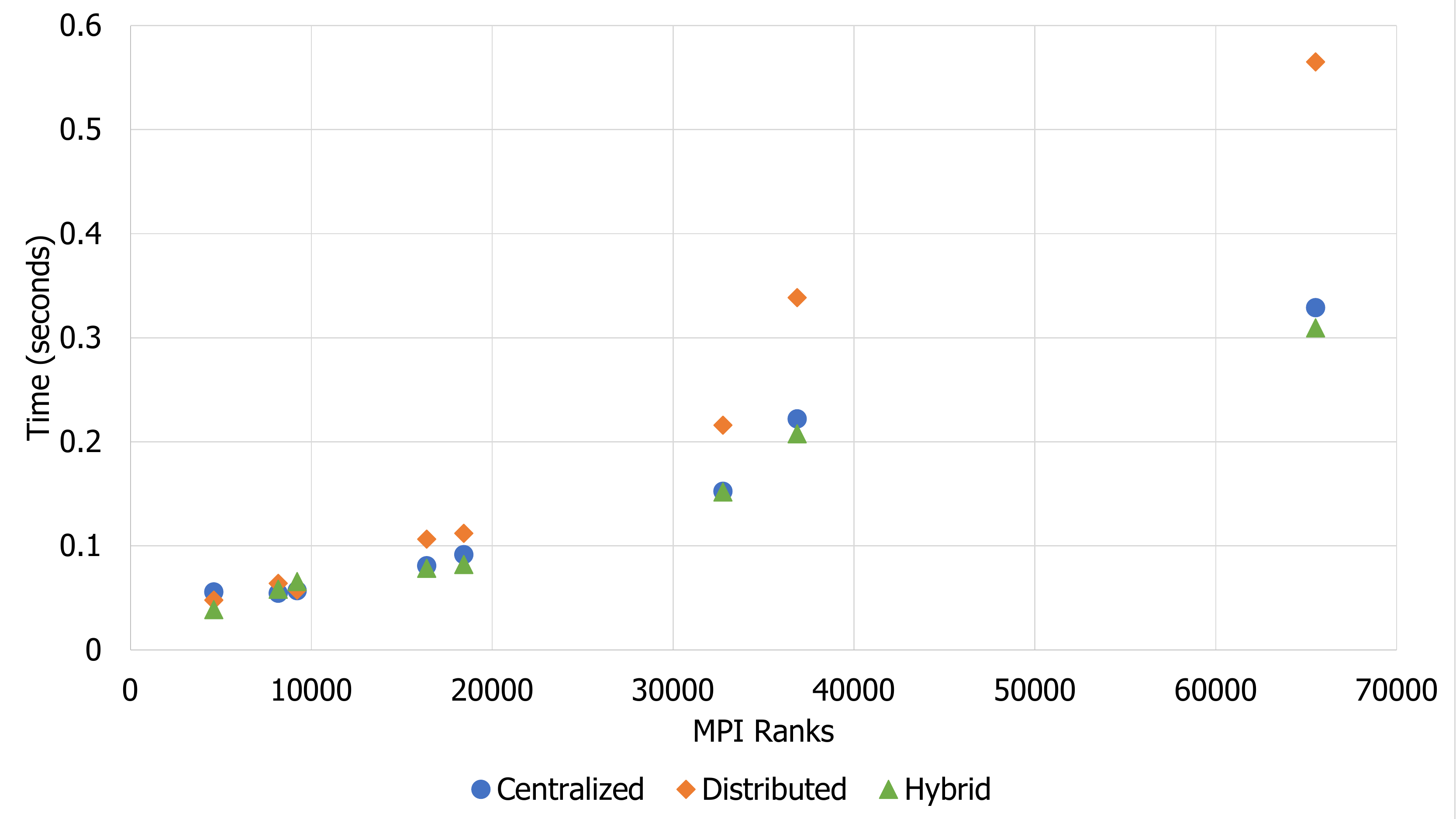} 
\vspace{-0.3cm}
\caption{On Quartz, total overhead for a load balancing step for centralized, distributed and hybrid algorithms. It is a weak scaling with $4$ elements per MPI rank, $5\times5\times5$ grid points per element, and about $343$ particles per element. The actual overhead expressed as number of time steps for $65,520$ MPI ranks was $1.94$ for the centralized, $3.35$ for the distributed, and $1.82$ for the hybrid algorithm.}
\label{fig:qtotoverhead}
\vspace{-0.4cm}
\end{figure}

\begin{figure}
\centering
\includegraphics[width=0.9\columnwidth]{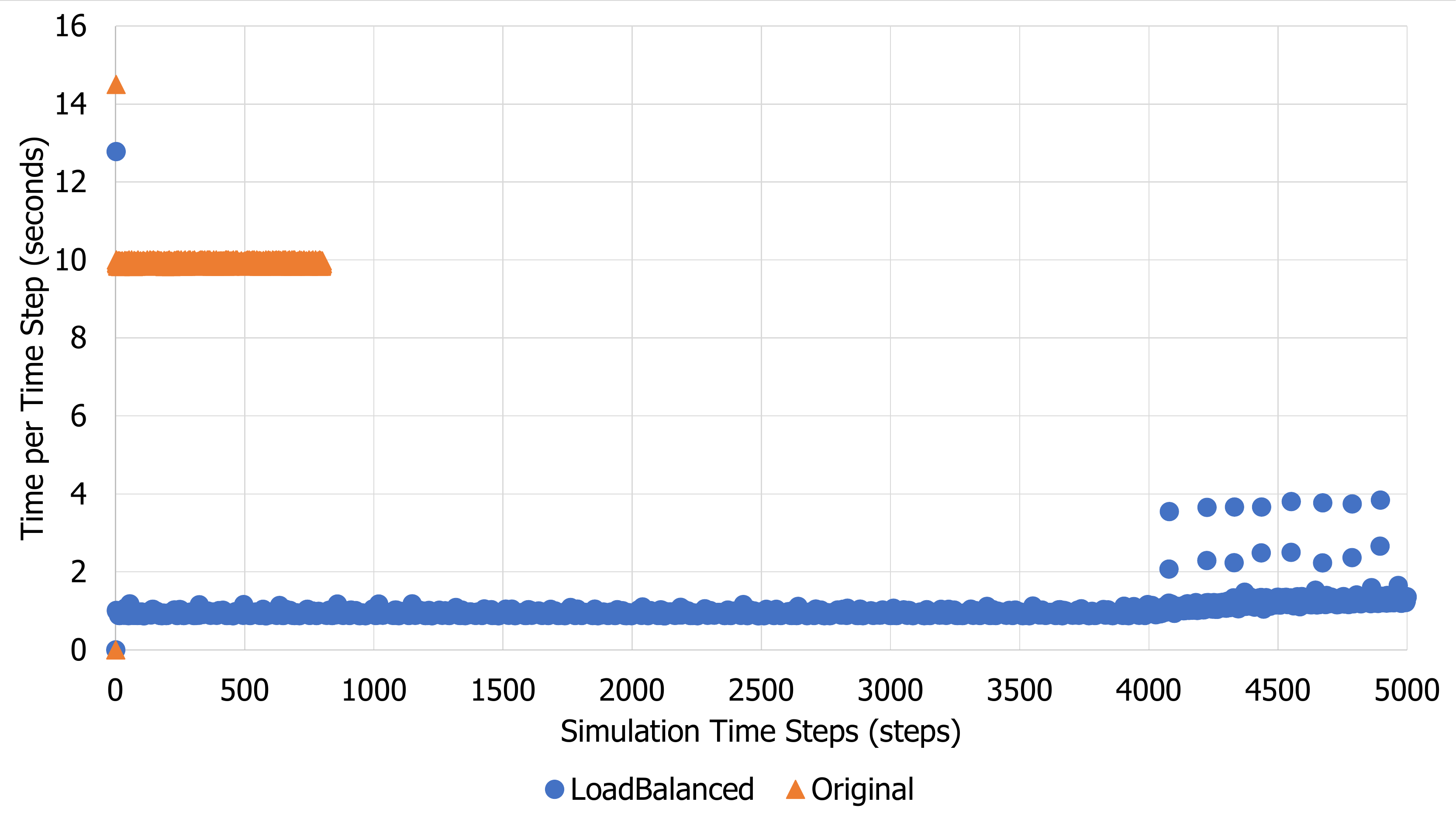} 
\vspace{-0.3cm}
\caption{Performance comparison between load-balanced and original versions of CMT-nek on Quartz. They were run on $67,206$ MPI ranks, that is $1,867$ nodes with $36$ cores per node. Adaptive hybrid load balancing was used. The average time per time step taken by the original version and the load balanced version were $9.92$ and $0.995$ seconds, respectively, giving us an overall speed-up factor of $9.97$. Original version did not finish in $2.2$ hours. 
}
\vspace{-0.3cm}
\label{fig:compareLBOrigQuartz}
\end{figure}

\vspace{-0.3cm}
\subsection{Experiments on Quartz}\label{subsec:expQuartz}
Figure \ref{fig:qtotoverhead} shows the overhead of the centralized, distributed and hybrid load-balancing algorithms on Quartz. It is a weak scaling with $4$ elements per MPI rank and about $343$ particles on an average per element. The variable $lelt$ was set to $16$. Each spectral element consists of $5\times 5\times 5$ grid points. The overhead includes time taken for each of the following steps: 1) remapping elements to processors; 2) packing, sending, and unpacking received elements and particles; and 3) reinitialization of data structures that are used in computation. The horizontal axis represents the number of MPI ranks while the vertical axis represents the time in seconds taken to load balance the application. 

The overhead incurred by a load-balancing step increases with the number of MPI ranks. Ideally, the distributed algorithm should take less time than the centralized algorithm with  increasing MPI ranks since there is no processor $P_0$ bottleneck in it. However, on Quartz the centralized algorithm is faster due to a higher ratio of communication-time to computation-time on the system and the distributed algorithm is rich in communication especially in $MPI\_ALLGATHERV$. The hybrid algorithm, eliminates calls to $MPI\_ALLGATHERV$, as well as, the part in the centralized algorithm where all processors send their element loads to $P_0$. As we can see from Figure \ref{fig:qtotoverhead}, the hybrid algorithm was the fastest. The actual overhead for $65,520$ MPI ranks for centralized, distributed and hybrid was $0.33$, $0.57$ and $0.31$ seconds, respectively. Compared to the time per time step which was $0.17$ seconds, the overhead expressed as a number of time steps was $1.94$ for the centralized, $3.35$ for the distributed, and $1.82$ for the hybrid algorithm. This makes dynamic load balancing practical for a large class of simulations. For these experiments, the total number of time steps was 100, and load-balancing took place every 10 steps.
Thus, we found that the overhead for load balancing is low and scales very well with the number of processors.

The load-balanced and non-load-balanced (original) codes were run on $67,206$ MPI ranks on Quartz, that is $1,867$ nodes with $36$ cores per node. The grid size per element was $5\times 5\times 5$ and the total number of elements was $900,000$. The variable $lelt$ was set to $120$ elements. The total number of particles was $1.125\times 10^{9}$, obtained as $1250$ particles per element on an average. Initially, the percent of elements that have particles is 6.1\% of the total number of elements. 

Figure~\ref{fig:compareLBOrigQuartz} compares a trace of the CPU time taken per simulation time step for load-balanced versus the original. Adaptive hybrid load balancing was used in this example. The average time per time step for the original and the load balanced versions were $9.92$ and $0.995$ seconds, respectively. Thus, we gained an overall speed-up of $9.97$ using load balancing algorithm. During the duration of simulation, apart from the compulsory load balancing that happens before simulation time step $1$, CMT-nek begins load balanced at $4,077$ time steps and after that, it will automatically load balance by itself (the small blue dot there represents the time taken by next step after load balance). Note that, in a production run, CMT-nek simulation will run for several hundred thousand time steps (each time step is less than a micro second of real time), hence first automatic rebalancing happening at time step $4,077$ is not too late in that context. Original version did not finish in $2.2$ hours. 

\begin{figure}
\centering
\includegraphics[width=0.9\columnwidth]{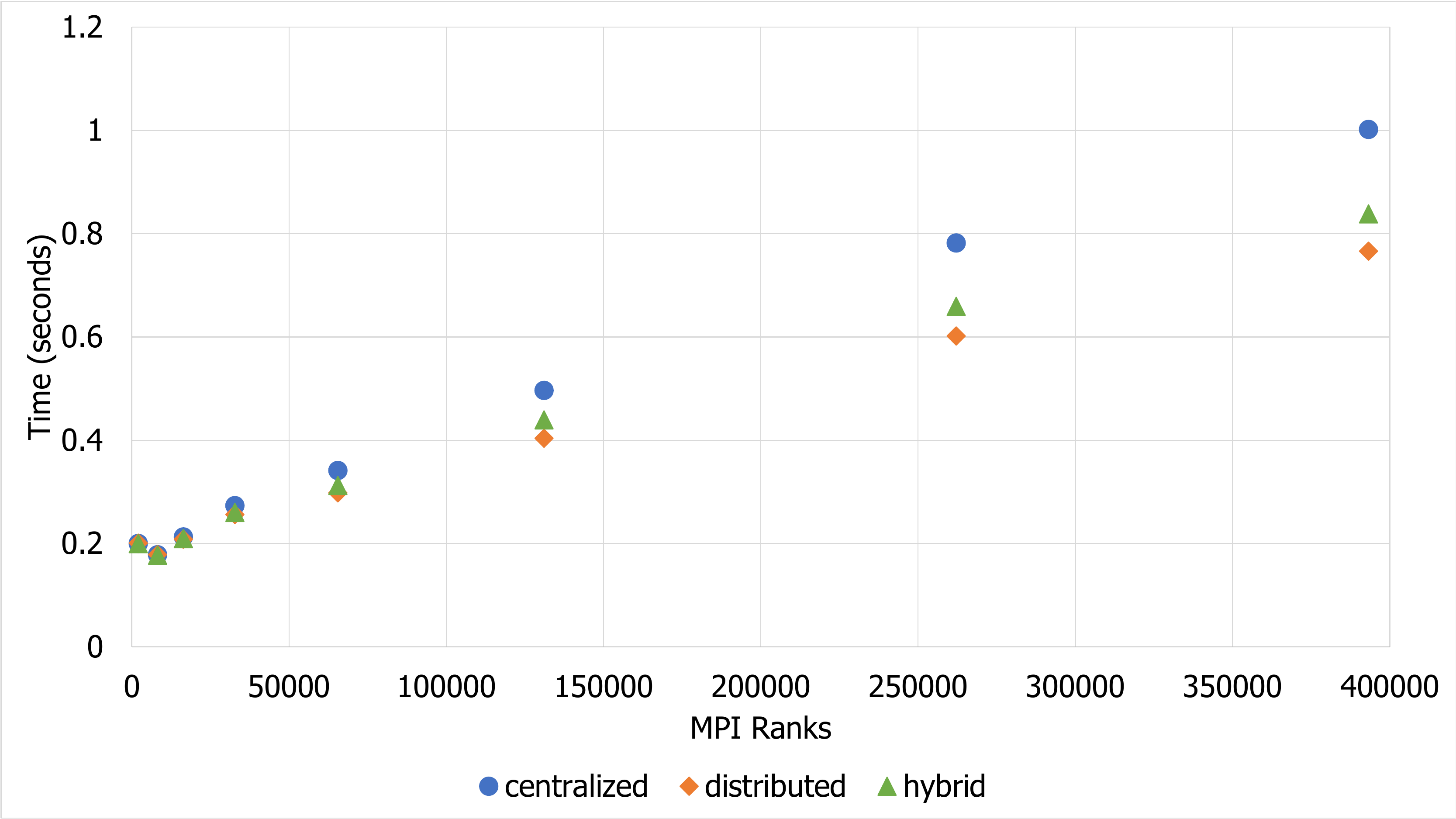}
\vspace{-0.3cm}
\caption{On Vulcan, total overhead for a load-balancing step for centralized, distributed and hybrid algorithms. It is a weak scaling with $2$ elements per MPI rank, $5\times5\times5$ grid points per element, and about $343$ particles per element. The actual overhead expressed as the number of time steps for $393,216$ MPI ranks was $3.03$ for the centralized, $2.33$ for the distributed, and $2.55$ for the hybrid algorithm.}
\vspace{-0.3cm}
\label{fig:totoverhead}
\end{figure}

\vspace{-0.3cm}
\subsection{Experiments on Vulcan}
We now evaluate the load-balancing algorithms on Vulcan. Figure~\ref{fig:totoverhead} shows the total overhead for a load-balancing step using the centralized, distributed and hybrid load-balancing algorithms. It is a weak scaling study so problem size increases proportionally to the number of MPI ranks, that is $2$ elements per MPI rank and $343$ particles per element on average. As we can see from Figure~\ref{fig:totoverhead}, the load-balancing overhead increases with an increasing number of total MPI ranks. Especially, distributed algorithm was faster than the centralized and hybrid algorithms. That is because of a lower ratio of communication-time to computation-time on this platform.
The actual overhead for $393,216$ MPI ranks for centralized, distributed and hybrid algorithm was $1.00$, $0.77$ and $0.84$ seconds respectively. Compared to the time per time step which was $0.33$ seconds, the overhead expressed as the number of time steps was $3.03$ time steps for the centralized, $2.33$ for the distributed, and $2.55$ for the hybrid algorithm. The variable $lelt$, which is the maximum number of elements on an MPI rank, was set to $8$ for these overhead runs. Of these experiments, the total number of time step was 100 and load balancing took place every 10 steps. Again, we can see from these results that the overhead for load balancing is low and scales very well with the number of processors.

The load-balanced and original codes were run on $65,536$ MPI ranks, that is $16,384$ nodes with $4$ cores per node. The grid size per element was $5\times 5\times 5$ and the total number of elements was $900,000$. The total number of particles was $1.125\times 10^{9}$ obtained as $1250$ particles per element on average. Initially, the percent of elements that have particles was 6.1\% of the total number of elements. The variable $lelt$ was set to $140$ for the load-balanced version.

\begin{figure}
\centering
\includegraphics[width=0.9\columnwidth]{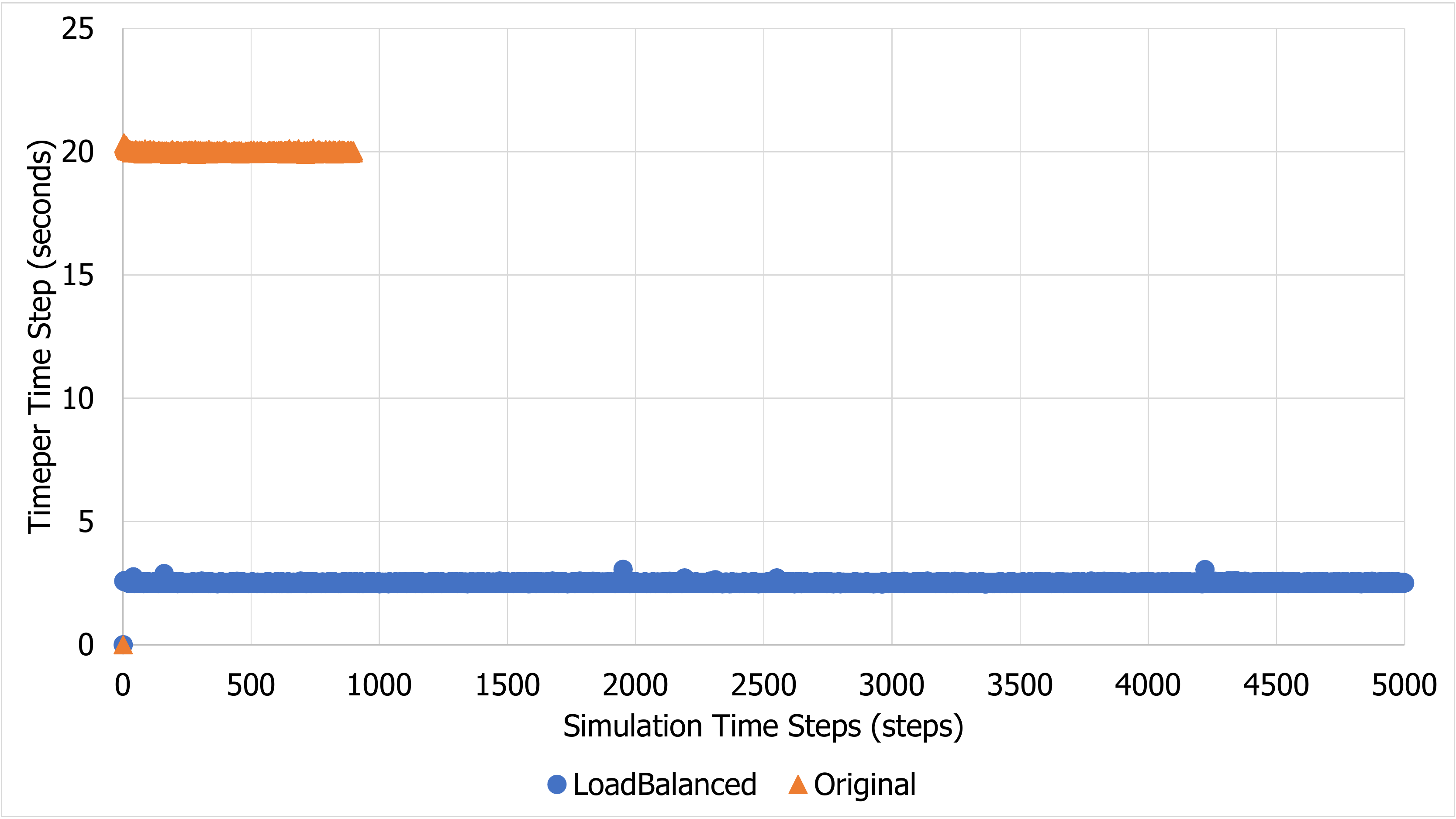} 
\vspace{-0.3cm}
\caption{Performance comparison between load-balanced and original versions of CMT-nek on Vulcan. They were run on $65,536$ MPI ranks, that is $16,384$ nodes with $4$ cores per node. Adaptive distributed load balancing was used. The average time per time step for the original and the load-balanced versions was $20.00$ and $2.52$ seconds, respectively, giving us an overall speed-up of $7.9$. Original code did not finish in $5$ hours. 
}
\label{fig:lbvsOrig_Vulcan}
\vspace{-0.3cm}
\end{figure}

\begin{figure}
\centering
\includegraphics[width=0.9\columnwidth]{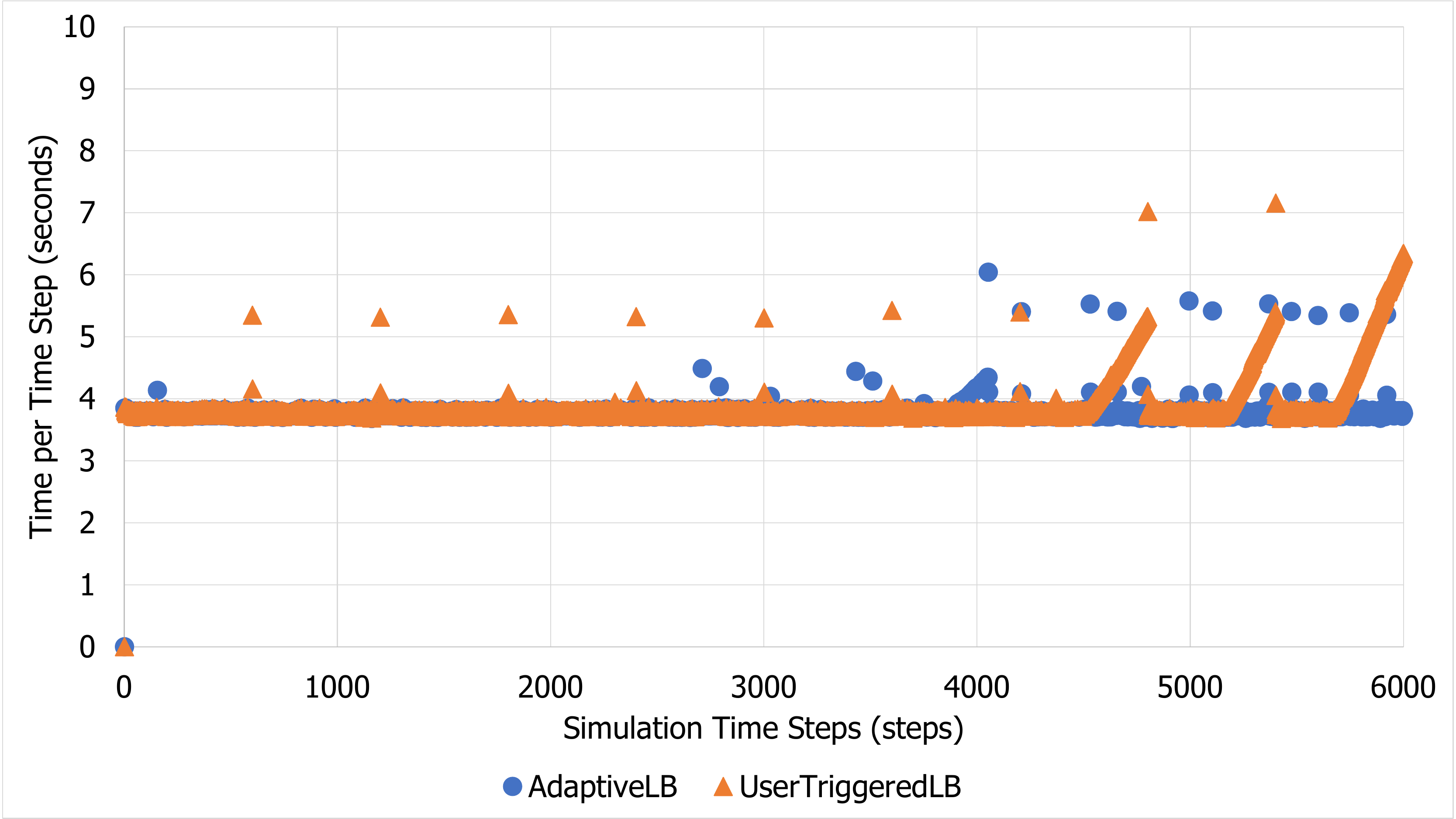} 
\vspace{-0.3cm}
\caption{Performance comparison between adaptive load-balanced and user-triggered load balanced versions of CMT-nek on Vulcan. They were run on $32,768$ MPI ranks, that is $8,192$ nodes with $4$ cores per node. Distributed load-balancing algorithm was used. The time per time step for the user-triggered and the adaptive load-balanced versions was $4.17$ and $3.78$ seconds, respectively, for the last $2,000$ steps, giving us an overall improvement of $9.4\%$. 
}
\label{fig:adplbvsUTLB_Vulcan}
\vspace{-0.3cm}
\end{figure}

Figure~\ref{fig:lbvsOrig_Vulcan} shows the differences in performance of load-balanced versus the original CMT-nek on Vulcan. The original version didn't finish in 5 hours. The time per time step for the original and the load-balanced versions was $20.00$ and $2.52$ seconds, respectively, giving us an overall speed-up of $7.9$. Load balancing happened before simulation time step 1. There was no need to load balance after that since the time per time step didn't increase over the threshold set to trigger load balancing.

Figure \ref{fig:adplbvsUTLB_Vulcan} shows a comparison between the adaptive load-balancing and user-triggered load-balancing algorithms. For the user-triggered load-balancing algorithm, the $k=500$, thus load balance is triggered  every $500$ time steps. As we can see from the figure, there is no performance degradation  in the first $4,000$ time steps, making any load balancing redundant during this time. However, right after step $4,000$ performance degrades sharply, requiring frequent load-balancing. The user-triggered load-balancing algorithm is insensitive to these performance variations and continues to load balance every $500$ time steps. The average time per time step from step $4,000$ to step $6,000$ taken by the adaptive and user-triggered load-balancing versions was $3.78$ and $4.17$ seconds, respectively.  Thus, adaptive load-balancing algorithm gained an overall improvement of $9.4$\% compared to the user specified triggered load-balancing algorithm, and further the load balancing happens automatically without requiring any intervention by the user.

\vspace{-0.3cm}
\section{Conclusion\label{sec:conclusion}}
In this paper, we have shown that an architecture-independent mapping that reorders the spectral elements (and corresponding particles), followed by mapping and remapping the ordered spectral elements is an effective  strategy for load balancing CMT-nek, a compressible, multiphase turbulent flow application. We have developed centralized, distributed and hybrid load-balancing algorithms and have studied their comparative overheads on two platforms, namely, Intel Broadwell and IBM BG/Q. Our results show that using this approach, we can potentially scale the application up to several million cores and still achieve reasonable load-balancing overhead and overall load balance. Using our load-balanced code we obtained a speed-up factor of up to $9.97$. Given that scientific applications run simulations for weeks, this speed-up is significant since it will not only save valuable computing resources, but due to smaller execution times, will also result in fewer runtime failures. We also presented a simple adaptive load-balancing strategy which  automatically triggers load balancing when needed. Compared to the user triggered load balancing, the adaptive load balancing strategy can additionally save time up to $9.4\%$. 
The current work was focused on one-dimensional expansion of particles. For many simulations, this expansion happens in two (cylindrical) or three (spherical) dimensions. We believe that improvements using our load-balancing schemes will be even larger than for the one-dimensional expansion reported in this paper.

\eatme{
[6] I. Al-Furajh, S. Aluru, S. Goil, S. Ranka
Parallel construction of multidimensional binary search trees
IEEE Transactions on Parallel and Distributed Systems, 11 (2) (2000), pp. 136-148.

[7] S. Ranka, R.V. Shankar, K.A. Alsabti
Many-to-many personalized communication with bounded traffic
Proceedings of Frontiers of Massively Parallel Computation (1995).

[8] A. Choudhary, G. Fox, S. Hiranandani, K. Kennedy, C. Koelbel, S. Ranka,;1; Software support for irregular and loosely synchronous problems, Computing Systems in Engineering 3 (1-4), 43-52.}


\vspace{-0.2cm}
\section*{Acknowledgment}
This work was funded by the U.S. Department of Energy, National Nuclear Security Administration, Advanced Simulation and Computing Program, as a Cooperative Agreement under the Predictive Science Academic Alliance Program, Contract No. DOE-NA0002378. This material is based upon work supported by the National Science Foundation Graduate Research Fellowship under Grant No. DGE- 1315138.


\bibliographystyle{unsrt}
\bibliography{reference}

\begin{thebibliography}{10}

\bibitem{Banerjee-7839682}
T.~Banerjee, J.~Hackl, M.~Shringarpure, T.~Islam, S.~Balachandar, T.~Jackson,
  and S.~Ranka.
\newblock Cmt-bone—a proxy application for compressible multiphase turbulent
  flows.
\newblock In {\em High Performance Computing (HiPC), 2016 IEEE 23rd
  International Conference on}, pages 173--182. IEEE, 2016.

\bibitem{c19}
H.M. Tufo and P.F. Fischer.
\newblock Terascale spectral element algorithms and implementations.
\newblock In {\em In Proceedings of the 1999 ACM/IEEE conference on
  Supercomputing}, page~68. ACM, 1999.

\bibitem{nek5000}
P.~F. Fischer, J.~Lottes, S.~Kerkemeier, K.~Heisey, A.~Obabko, O.~Marin, and
  E.~Merzari.
\newblock http://nek5000.mcs.anl.gov, 2014.

\bibitem{dfm02}
M.~Deville, P.~Fischer, and E.~Mund.
\newblock {\em High-order methods for incompressible fluid flow}, volume~9.
\newblock Cambridge University Press, 2002.

\bibitem{hendrickson1993multidimensional}
B.~Hendrickson and R.~Leland.
\newblock Multidimensional spectral load balancing.
\newblock Technical report, Sandia National Labs., Albuquerque, NM (United
  States), 1993.

\bibitem{choudhary1992software}
A.~Choudhary, G.~Fox, S.~Hiranandani, K.~Kennedy, C.~Koelbel, S.~Ranka, and
  J.~Saltz.
\newblock Software support for irregular and loosely synchronous problems.
\newblock {\em Computing Systems in Engineering}, 3(1-4):43--52, 1992.

\bibitem{lieber2017highly}
M.~Lieber and W.~Nagel.
\newblock Highly scalable sfc-based dynamic load balancing and its application
  to atmospheric modeling.
\newblock {\em Future Generation Computer Systems}, 2017.

\bibitem{menon2012automated}
H.~Menon, N.~Jain, G.~Zheng, and L.~Kale.
\newblock Automated load balancing invocation based on application
  characteristics.
\newblock In {\em Cluster Computing (CLUSTER), 2012 IEEE International
  Conference on}, pages 373--381. IEEE, 2012.

\bibitem{surmin2015dynamic}
I.~Surmin, A.~Bashinov, S.~Bastrakov, E.~Efimenko, A.~Gonoskov, and I.~Meyerov.
\newblock Dynamic load balancing based on rectilinear partitioning in
  particle-in-cell plasma simulation.
\newblock In {\em International Conference on Parallel Computing Technologies},
  pages 107--119. Springer, 2015.

\bibitem{ferraro1993dynamic}
R.~Ferraro, P.~Liewer, and V.~Decyk.
\newblock Dynamic load balancing for a 2d concurrent plasma pic code.
\newblock {\em Journal of computational physics}, 109(2):329--341, 1993.

\bibitem{plimpton2003load}
S.~Plimpton, D.~Seidel, M.~Pasik, R.~Coats, and G.~Montry.
\newblock A load-balancing algorithm for a parallel electromagnetic
  particle-in-cell code.
\newblock {\em Computer physics communications}, 152(3):227--241, 2003.

\bibitem{germaschewski2013plasma}
K.~Germaschewski, W.~Fox, S.~Abbott, N.~Ahmadi, K.~Maynard, L.~Wang, H.~Ruhl,
  and A.~Bhattacharjee.
\newblock The plasma simulation code: A modern particle-in-cell code with
  load-balancing and gpu support.
\newblock {\em arXiv preprint arXiv:1310.7866}, 2013.

\bibitem{nakashima2009ohhelp}
H.~Nakashima, Y.~Miyake, H.~Usui, and Y.~Omura.
\newblock Ohhelp: a scalable domain-decomposing dynamic load balancing for
  particle-in-cell simulations.
\newblock In {\em Proceedings of the 23rd international conference on
  Supercomputing}, pages 90--99. ACM, 2009.

\bibitem{pearce2014load}
O.~Pearce, T.~Gamblin, B.~De~Supinski, T.~Arsenlis, and N.~Amato.
\newblock Load balancing n-body simulations with highly non-uniform density.
\newblock In {\em Proceedings of the 28th ACM international conference on
  Supercomputing}, pages 113--122. ACM, 2014.

\bibitem{bhatele2009dynamic}
A.~Bhatel{\'e}, L.~Kal{\'e}, and S.~Kumar.
\newblock Dynamic topology aware load balancing algorithms for molecular
  dynamics applications.
\newblock In {\em Proceedings of the 23rd international conference on
  Supercomputing}, pages 110--116. ACM, 2009.

\bibitem{al2000parallel}
I.~Al-Furajh, S.~Aluru, S.~Goil, and S.~Ranka.
\newblock Parallel construction of multidimensional binary search trees.
\newblock {\em IEEE Transactions on Parallel and Distributed Systems},
  11(2):136--148, 2000.

\bibitem{ranka1995many}
S.~Ranka, R.~Shankar, and K.~Alsabti.
\newblock Many-to-many personalized communication with bounded traffic.
\newblock In {\em Frontiers of Massively Parallel Computation, 1995.
  Proceedings. Frontiers' 95., Fifth Symposium on the}, pages 20--27. IEEE,
  1995.

\bibitem{HacklCMTCAF}
J.~Hackl, M.~Shringarpure, R.~Koneru, M.~Delchini, and S.~Balachandar.
\newblock {A shock capturing discontinuous Galerkin spectral element method for
  curved geometry using entropy viscosity}.
\newblock {\em Computers and Fluids}.

\bibitem{TVDRK}
S.~Gottlieb and C.-W. Shu.
\newblock {Total variation-diminishing Runge-Kutta schemes}.
\newblock {\em {Math. Comp.}}, 67:73--85, 1998.

\bibitem{berrut2004}
J.~P. Berrut and L.~N. Trefethen.
\newblock Barycentric lagrange interpolation.
\newblock {\em SIAM review}, 46(3):501--517, 2004.

\bibitem{Ou:1997:PRA:264305.264312}
C.~W. Ou and S.~Ranka.
\newblock Parallel remapping of adaptive problems.
\newblock {\em J. Parallel Distrib. Comput.}, 42(2):109--121, May 1997.

\bibitem{Ou1996}
C.~W. Ou, S.~Ranka, and G.~Fox.
\newblock Fast and parallel mapping algorithms for irregular problems.
\newblock {\em The Journal of Supercomputing}, 10(2):119--140, Jun 1996.

\bibitem{Ou:1995:ALT:224538.224573}
C.~W. Ou, M.~Gunwani, and S.~Ranka.
\newblock Architecture-independent locality-improving transformations of
  computational graphs embedded in k-dimensions.
\newblock In {\em Proceedings of the 9th International Conference on
  Supercomputing}, ICS '95, pages 289--298, New York, NY, USA, 1995. ACM.

\bibitem{Tufo:2001:FPD:372836.372845}
H.~M. Tufo and P.~F. Fischer.
\newblock Fast parallel direct solvers for coarse grid problems.
\newblock {\em J. Parallel Distrib. Comput.}, 61(2):151--177, February 2001.

\bibitem{gapaper}
T.~Banerjee and S.~Ranka.
\newblock A genetic algorithm based autotuning approach for performance and
  energy optimization.
\newblock In {\em Green Computing Conference and Sustainable Computing
  Conference (IGSC), 2015 Sixth International}, pages 1--8. IEEE, 2015.

\end{thebibliography}
\end{document}